\documentclass[onecolumn]{aa}  
\usepackage{graphicx}
\usepackage{natbib}
\usepackage{txfonts}
\begin{document}
\title{New two-colour light curves of Q0957+561:\\  
       time delays and the origin of intrinsic variations}


\author{V. N. Shalyapin\inst{1}, L. J. Goicoechea\inst{2}, E. Koptelova\inst{3,4}, 
        A. Ull\'an\inst{5} \and R. Gil-Merino\inst{6}}

\institute{Institute for Radiophysics and Electronics, National Academy of Sciences of Ukraine, 
             12 Proskura St., Kharkov 61085, Ukraine\\
             \email{vshal@ire.kharkov.ua}
\and
             Departamento de F\'{\i}sica Moderna, Universidad de Cantabria, 
	       Avda. de Los Castros s/n, 39005 Santander, Spain\\
             \email{goicol@unican.es}
\and
             Sternberg Astronomical Institute, Universitetski pr. 13, 119992 Moscow, Russia\\
             \email{koptelova@xray.sai.msu.ru}
\and
		 Graduate Institute of Astronomy, Jhongli City, Taoyuan County 320, Taiwan\\
             \email{koptelova@astro.ncu.edu.tw}
\and
             Robotic Telescopes Group, Centro de Astrobiolog\'{\i}a (CSIC-INTA), associated to the 
             NASA Astrobiology Institute, Ctra de Ajalvir, km 4, 28850 Torrej\'on de Ardoz, Madrid, 
             Spain\\
             \email{ullanna@inta.es}
\and
             Instituto de F\'{\i}sica de Cantabria (CSIC-UC), Avda. de Los Castros s/n, 39005 
             Santander, Spain\\
             \email{gilmerino@ifca.unican.es}}

\titlerunning{New two-colour light curves of Q0957+561}
\authorrunning{Shalyapin et al.}


\abstract
{}   
{We extend the $gr$-band time coverage of the gravitationally lensed double quasar Q0957+561. New $gr$ light 
curves permit us to detect significant intrinsic fluctuations, to determine new time delays, and thus to 
gain perspective on the mechanism of intrinsic variability in Q0957+561.}
{We use new optical frames of Q0957+561 in the $g$ and $r$ passbands from January 2005 to July 2007. These
frames are part of an ongoing long-term monitoring with the Liverpool robotic telescope. We also introduce 
two photometric pipelines that are applied to the new $gr$ frames of Q0957+561. The transformation pipeline 
incorporates zero-point, colour, and inhomogeneity corrections to the instrumental magnitudes, so final 
photometry to the 1-2\% level is achieved for both quasar components. The two-colour final records are then 
used to measure time delays.}
{The $gr$ light curves of Q0957+561 show several prominent events and gradients, and some of them (in the 
$g$ band) lead to a time delay between components $\Delta t_{BA}$ = 417 $\pm$ 2 d (1$\sigma$). We do not 
find evidence of extrinsic variability in the light curves of Q0957+561. We also explore the possibility 
of a delay between a large event in the $g$ band and the corresponding event in the $r$ band. The 
$gr$ cross-correlation reveals a time lag $\Delta t_{rg}$ = 4.0 $\pm$ 2.0 d (1$\sigma$; the $g$-band event 
is leading) that confirms a previous claim of the existence of a delay between the $g$ and $r$ band in this 
lensed quasar.}
{The time delays (between quasar components and between optical bands) from the new records and previous 
ones in similar bands indicate that most observed variations in Q0957+561 (amplitudes of $\sim$ 100 mmag and
timescales of $\sim$ 100 d) are very probably due to reverberation within the gas disc around the 
supermassive black hole.}


\maketitle


\section{Introduction}

Studies of optical continuum variability in gravitationally lensed quasars (GLQs) have a main
advantage: one is usually able to disentangle intrinsic from extrinsic signal in GLQs 
\citep[e.g.,][Paper I]{Kun97,Par06,Goi08}. Following the original idea by \citet{Ref64}, 
intrinsic variations in brightness records of GLQs have mainly been used to estimate global time 
delays between components, and to discuss the structure of galaxy mass halos and the expansion 
rate of the Universe \citep[e.g.,][and references therein]{Koc04a}. Less effort has been devoted 
to investigating the nature of intrinsic fluctuations, which are generated by mechanisms of 
variability in lensed quasars. This can be done by measuring time delays between components 
and between optical bands, using prominent events in segments of long-term light curves. Time 
delays between two given components of a GLQ (determined from different pairs of twin features) 
arise from gravitational lensing of flares in the variable source. While the flaring of a 
well-defined emission region (e.g., a ring of the accretion disc) produces a set of similar 
delays, the existence of flares in some widely separated zones can lead to important time 
delay differences \citep{Yon99}. For either of the two components, time delays between optical 
bands (or interband time delays) refer to time lags arising from physical phenomena within the 
quasar. 

The gravitationally lensed double quasar \object{Q0957+561} at $z$ = 1.41 \citep{Wal79} has been
monitored photometrically in different optical bands with different telescopes. To derive a 
global time delay between quasar components, some previous studies used large data sets  
incorporating all kinds of fluctuations, i.e., noisy or poorly sampled features as well as 
noticeable gradients and events on several timescales. These large data sets are based on frames 
that were taken in the 1980s and 90s, and they lead to a global delay of about 423$-$425 d 
\citep{Osc01,Ova03a}. The Apache Point Observatory (APO) experiment permitted investigators to 
follow-up the variability in the $g$ and $r$ bands during the 1995 and 1996 seasons, i.e., 
covering 1.5 years \citep{Kun97}. This monitoring programme with the APO 3.5 m telescope produced 
accurate light curves of both components \object{Q0957+561A} and \object{Q0957+561B}, which show 
sharp intrinsic features with high signal-to-noise ratio ($S/N \geq$ 3). From the APO main twin 
events in the $g$ band (a prominent event in A and the replica event in B; $S/N \sim$ 6.5), 
\citet{Kun97} also reported a gravitational lens time delay of 417 $^{+3}_{-4}$ d (95\% 
confidence interval). Complementary to this result, \citet{Col01} found that the $r$-band main 
twin events lag with respect to the ones in the $g$-band by 3.4 $^{+1.5}_{-1.4}$ d (68\% 
confidence interval), and this interband delay was interpreted as clear evidence for reprocessing 
in the accretion disc of the quasar.  

\citet{Goi02} reanalysed the APO $g$-band data set to obtain two different gravitational lens 
time delays of 417.0 $\pm$ 0.6 d (68\% confidence interval) and 432.0 $\pm$ 1.9 d (68\% 
confidence interval) depending on the features taken as reference. The longest delay corresponds 
to the APO secondary twin events ($S/N \sim$ 3) and it clearly disagrees with the 417-day value. 
The APO main and secondary twin events in the $g$ band are associated with a main flare and 
a secondary flare in the variable source, respectively. From the time delay difference of 15 
$\pm$ 2 days (68\% confidence interval) and using standard cosmological parameters ($\Omega$ = 
0.3, $\Lambda$ = 0.7), one can also determine a minimum size for the variable source (minimum
distance between both flares) of 300 pc \citep{Yon99,Goi02,Yon03}. Several physical sources are 
consistent with this spatial constraint (from multiple gravitational lens delays) and the 
interband delay for the main twin events, but either a nuclear accretion disc plus a 
circumnuclear stellar region or a nuclear accretion disc plus an optical jet are the most 
probable ones \citep[e.g.,][]{Hut03}. We note that the 424-d global time delay between 
components \citep{Osc01,Ova03a} coincides with the average of both APO gravitational lens time 
delays. This suggests the presence of two or more gravitational lens delays in the current large 
data sets \citep[for alternative explanations, see][]{Sch05,Hir07}. 

In the present study, we substantially extend the $gr$-band time coverage of 
\object{Q0957+561}. The key idea is to use new $gr$ light curves to detect prominent intrinsic 
events similar to the APO ones. These new features should allow us to determine new time delays 
and to improve our understanding of the mechanism causing the intrinsic variability. The paper 
is organised as follows: in Section 2, we present new data of \object{Q0957+561} based on recent 
observations with the Liverpool 2 m telescope (LT) in the $g$ and $r$ bands, spanning 2.5 years. 
We describe the observations, the pre-processing, and the photometric procedure for determining 
calibrated and corrected magnitudes of field stars and quasar components. This last reduction 
procedure consists of two new pipelines specially designed for the LT. In Section 3 we study the 
time delays between the two components of \object{Q0957+561} as well as the possible delays 
between the $g$ and $r$ band in the new data set. In Section 4 we summarize our results. 
From the APO and LT delays of \object{Q0957+561}, we also discuss the origin of the observed 
variations with an amplitude of $\sim$ 100 mmag and lasting $\sim$ 100 d.


\section{Data acquisition and reduction}

\subsection{Observations and pre-processing}

Liverpool Quasar Lens Monitoring (LQLM) I is the first phase of an optical follow-up of lensed
quasars, undertaken using the RATCam optical CCD camera on the Liverpool robotic telescope 
\citep{Ste04} between January 2005 and July 2007. The first scientific output of LQLM I was 
reported in Paper I, and we concentrate here on the observations of \object{Q0957+561} in the 
$g$ and $r$ filters. The field of view  and the pixel scale (binning 2$\times$2) were $\sim 
4\farcm6\times4\farcm6$ and $0\farcs278$, respectively. The exposure times were 100$-$200 s ($g$ 
band) and 120 s ($r$ band). We obtained 286 frames in the $g$ band and 264 frames in the $r$ 
band. The LT observed for a total science time of $\sim$ 22.6 h during the 2.5-year programme of 
\object{Q0957+561}.

A pre-processing pipeline is applied to all RATCam frames\footnote{See the Web site 
http://telescope.livjm.ac.uk/Info/TelInst/Inst/RATCam/index.php.}. This performs three basic 
instrumental reductions: bias subtraction, trimming of the overscan regions, and flat fielding. 
We also apply a bad-pixel mask \citep[made available by the Angstrom project;][]{Ker06}, and 
correct bad pixels on the CCD. The next step is the pre-selection of frames, based on individual 
inspection, to assure that exposures verify some elemental conditions (e.g., that the telescope 
pointing was accurate enough so that the lens system was included in the field of view, that 
there is no strongly degraded signal, etc), and that seeing ($FWHM$) and sky level (background) 
values do not exceed reasonable bounds. We only consider frames with $FWHM < 3\arcsec$ due to 
the separation between the two quasar components (A and B) of $\sim 6\arcsec$. The pre-selected 
database contains 199 frames in the $g$ band and 210 frames in the $r$ band. This means that 
$\sim$ 75\% of the original LT frames were initially useful.

\subsection{Instrumental photometry}

In a first step, we take a reference frame, i.e., a high-quality frame with small $FWHM$ and large 
signal-to-noise ratio ($SNR$). We then measure the positions (with respect to the left bottom corner 
of the reference frame) of seven reference stars and both quasar images. We select the 7 brightest 
stars in the Sloan Digital Sky Survey (SDSS) catalog\footnote{See the DR6 Catalogue Archive Server 
site http://cas.sdss.org/astrodr6/en/. Funding for the SDSS has been provided 
by the Alfred P. Sloan Foundation, the Participating Institutions, the NASA, the NSF, the U.S. 
Department of Energy, the Japanese Monbukagakusho, and the Max Planck Society. The SDSS is managed by 
the Astrophysical Research Consortium (ARC) for the Participating Institutions. The Participating 
Institutions are The University of Chicago, Fermilab, the Institute for Advanced Study, the Japan 
Participation Group, The Johns Hopkins University, Los Alamos National Laboratory, the 
Max-Planck-Institute for Astronomy (MPIA), the Max-Planck-Institute for Astrophysics (MPA), New Mexico 
State University, University of Pittsburgh, Princeton University, the United States Naval Observatory, 
and the University of Washington.} \citep[e.g.,][]{Ade07}. These stars, having $g$(SDSS) and $r$(SDSS) 
magnitudes below 18 and 17, respectively, were labeled as X, G, F, H, D, E, and R stars in Figure 1 
and Table 1 of \citet{Ova03a}. Several Image Reduction and Analysis Facility (IRAF)\footnote{IRAF is 
distributed by the National Optical Astronomy Observatory, which is operated by the Association of 
Universities for Research in Astronomy (AURA) under cooperative agreement with the National Science 
Foundation.} tasks are also used to identify the available reference objects (in general, less than 7 
stars) and the quasar components in the rest of the frames. 

In a second step, our photometric pipeline performs aperture photometry of bright field stars and 
quasar images. This IRAF procedure is used to estimate initial instrumental fluxes (sources and their 
associated backgrounds) and to improve the initial source positions on each frame. The pipeline also
cuts the original frames in order to produce square subframes with 64 pixels per side: the system 
subframe (around the centre of the lens system) and subframes of stars (around the bright stars), and 
makes a PSF subframe containing the clean 2D profile of the H star (removing the local background). 
This last empirical PSF is required when performing PSF fitting. The point-like sources (quasar 
components and stars) are modelled by means of the empirical PSF, whereas the extended source (lensing 
elliptical galaxy) is modelled by a de Vaucouleurs profile convolved with the empirical PSF. After 
obtaining all subframes for a given frame, PSF photometry on the stellar and system (crowed field) 
subframes is performed with IMFITFITS software \citep{McL98}. The pipeline is written in the Python 
programming language\footnote{See the Web site http://www.python.org/.}, and incorporates the 
capabilities of IRAF (through the PyRAF interface) and IMFITFITS, as well as additional numerical and 
graphical tools.

To determine accurate quasar fluxes, one needs to use a set of constraints. For \object{Q0957+561}, 
the most relevant constraints were obtained from Hubble Space Telescope (HST) frames 
\citep{Ber97,Kee98,Koc08}: positions of the B component and the lensing galaxy relative to the A 
component, and the optical structure of the galaxy, i.e., effective radius, ellipticity, and position 
angle (a de Vaucouleurs profile was fitted to HST images). Due to the relatively low brightness of 
the lensing galaxy in the frames and the proximity of the B component to the galaxy, we determine the 
galaxy-to-H star ratio ($GAL/H$) in the $gr$ bands from the best LT frames, in terms of $FWHM$ and 
$SNR$ values. The H star is relatively bright and it is present in all frames. We then apply the 
pipeline to all frames (whatever their qualities) in each optical filter, by setting the galaxy 
fluxes to those derived from the $GAL/H$ ratio and H star fluxes, and allowing the remaining free 
parameters to vary. 
                                               
\begin{figure}
\centering
\includegraphics[angle=0,width=\textwidth]{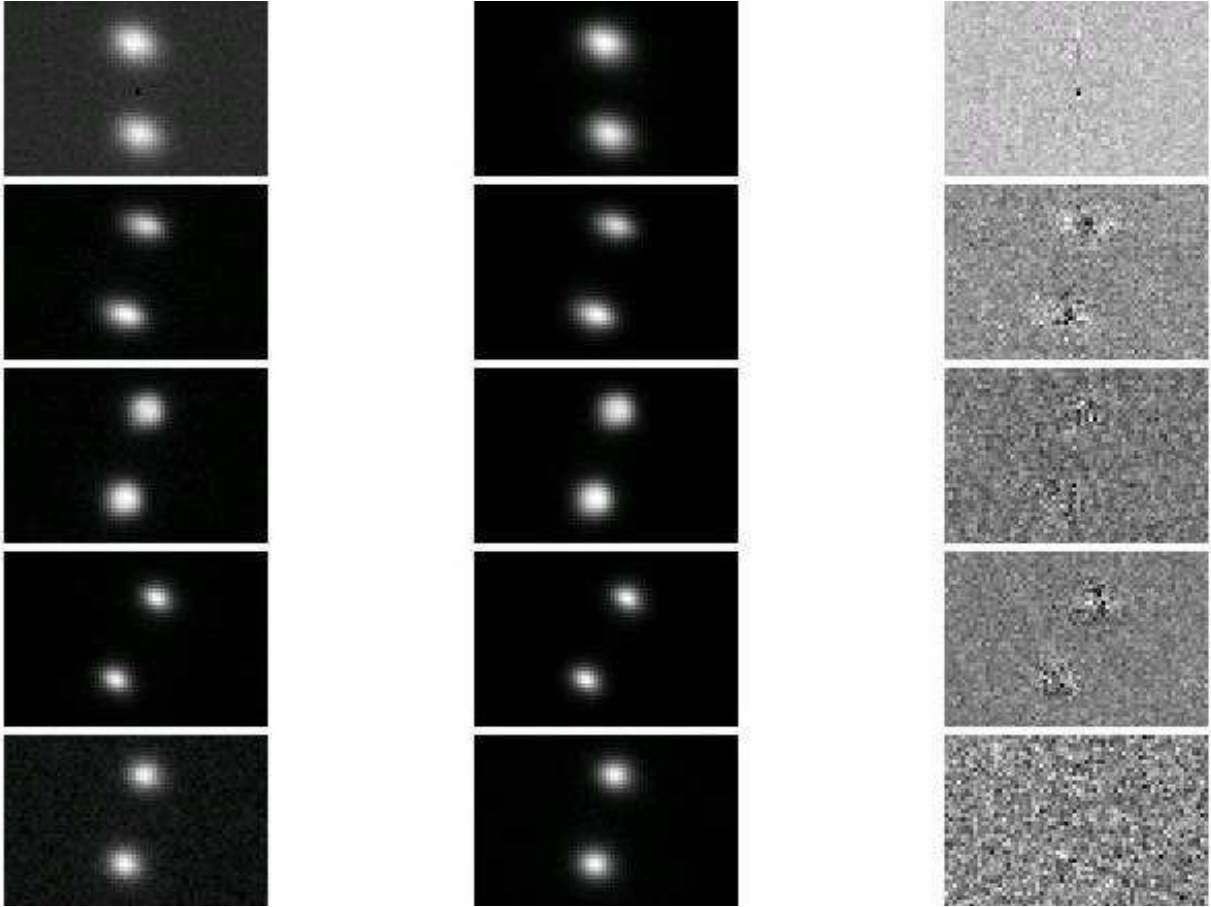}
\caption{Observations of Q0957+561 with the Liverpool robotic telescope in the $g$ band. We display 
system subframes (left panels), model subframes (middle panels), and residual subframes (right panels) 
for five frames taken during the 2.5-year monitoring period (see main text).}
\label{imfitfitsout}
\end{figure}

The photometry pipeline output includes the system subframes, their model subframes (best fits) and 
the associated residual subframes (system subframes after subtracting model subframes). In 
Fig.~\ref{imfitfitsout} we show system subframes (left panels), model subframes (middle panels), and 
residual subframes (right panels) corresponding to five frames in the $g$ band. From top to bottom: 
March 16, 2005 ($FWHM = 2\farcs03$, $SNR$ = 215, $\chi^2$ = 1.44), November 9, 2005 ($FWHM = 1\farcs75$, 
$SNR$ = 313, $\chi^2$ = 1.32), April 26, 2006 ($FWHM = 1\farcs76$, $SNR$ = 371, $\chi^2$ = 2.15), 
October 21, 2006 ($FWHM = 1\farcs48$, $SNR$ = 321, $\chi^2$ = 1.30), and May 28, 2007 ($FWHM = 
1\farcs63$, $SNR$ = 120, $\chi^2$ = 1.43), where the $SNR$ values are inferred from the A images 
(having fluxes similar to those of B images) and the $\chi^2$ values quantitatively describe the 
quality of the fits (i.e., these represent the standard reduced $\chi^2$ for the best fits). All 
subframes in Fig.~\ref{imfitfitsout} have been expanded by a factor of 2. The visual comparison between 
left and middle panels as well as the inspection of patterns of residual brightness (right panels) 
indicate that the photometric method works well. The pipeline also produces a basic data release file 
containing values of all relevant instrumental fluxes (stars and quasar images) and relative 
instrumental magnitudes of both quasar components, e.g., $g_A^* - g_E^*$ and $g_B^* - g_E^*$ in the $g$ 
band. To check the reliability of our PSF fitting procedure, we applied a deconvolution technique 
\citep{Kop05} to two sets of frames in October-December 2005 ($gr$ bands). The relative instrumental 
magnitudes from the deconvolution method agreed with the records from the PSF fitting technique 
\citep[see Fig. 3 of][]{Goi07}.  

\subsection{Calibrated and corrected magnitudes}

We use a transformation pipeline (in the Python programming language) to obtain SDSS magnitudes from 
instrumental magnitudes that are corrected for systematic effects. The whole calibration-correction 
process is outlined in Appendix A. Only frames with $SNR \geq$ 100 over \object{Q0957+561A} are taken 
into account (see however Section 3). In the $g$ band, this selection leads to 170 frames. In the $r$ 
band, besides the $SNR$ based selection, the surviving frames from the first season (January-June 
2005) are also removed. Several of these $\sim$ 10 $r$-band frames with $SNR$ above 100 (first season) 
are characterized by an anomalous image formation. Thus, the high-quality data set in the $r$ band 
includes 167 frames. 

\begin{figure}
\centering
\includegraphics[angle=-90,width=10cm]{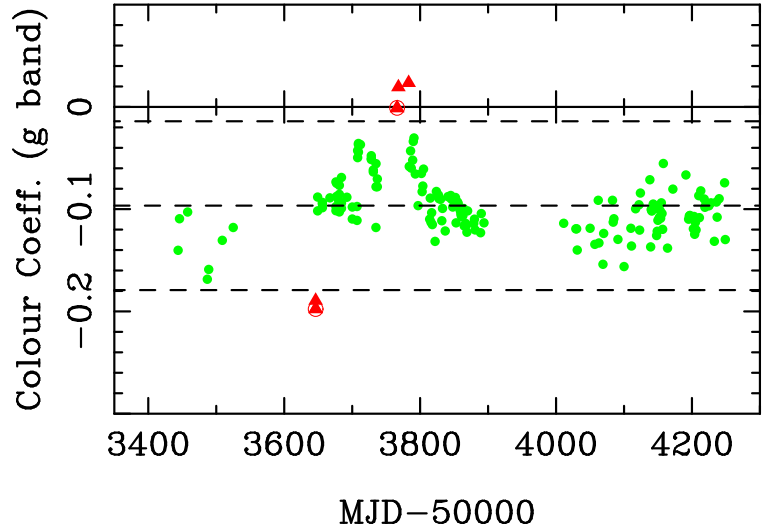}
\caption{Colour coefficient in the $g$ band. The values are distributed around the central discontinuous 
line (average coefficient), and most of them are placed between the top and bottom discontinuous lines
(filled circles). Only seven extreme values (triangles and open circles) exceed these limits.}
\label{colour}
\end{figure}
\begin{figure}
\centering
\includegraphics[angle=-90,width=7cm]{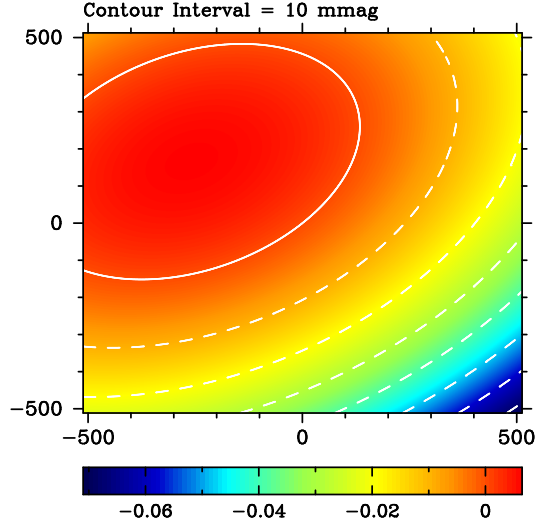}
\caption{Inhomogeneity map in the $g$ band. The zero inhomogeneity level is described by means of a 
continuous line that crosses the centre of the 1024$\times$1024 camera. Pixels inside this zero level 
contour have a positive inhomogeneity of a few mmag, whereas the rest of the pixels have a negative 
inhomogeneity that varies between a few mmag and tens of mmag. We explicitly show contours of $-$10, 
$-$20, $-$30, $-$40, $-$50, $-$60, and $-$70 mmag.}
\label{inhomomap}
\end{figure}

The transformation pipeline fits the deviations between instrumental and standard $g$ magnitudes of the 
7 reference stars to the transformation model that incorporates a zero-point term ($\alpha_g$), a colour 
coefficient ($\beta_g$), and inhomogeneity coefficients ($\gamma_{g, nm}$). Eq. (A.11) shows the 
relationship between the observed magnitude deviation and the model to describe it. The zero-point term 
and the colour coefficient are allowed to vary over time because the atmospheric and instrumental 
conditions significantly evolve during the 2.5 years of monitoring. The last ingredient of the model is 
a linear-quadratic inhomogeneity term, which is related to the 2D position on the CCD and tries to 
correct the possible inhomogeneous response over the camera area \citep[e.g.,][]{Man01,Mag04}. Each 
source occupies different positions on the CCD area during the robotic monitoring period, so this could 
complicate the collecting of accurate brightness records.

With respect to the least squares fit, in Fig.~\ref{colour} we plot the solution of $\beta_g$. The 
$\beta_g$ values are distributed around an average colour coefficient $\langle \beta_g \rangle = 
-0.097$ (central discontinuous line in Fig.~\ref{colour}), which is close to the typical coefficient 
(see comments in Appendix A). The scatter is $\sigma(\beta_g)$ = 0.033, and the  $\langle \beta_g 
\rangle \pm 2.5\sigma(\beta_g)$ limits also appear in Fig.~\ref{colour} (top and bottom discontinuous 
lines). In relation to the average coefficient, there are seven extreme values representing changes 
from 100\%, i.e., values around either $-$0.2 or 0.0 (triangles and open circles in Fig.~\ref{colour}). 
The first two triangles and the first open circle correspond to the first night after a realuminisation 
of the mirror and other maintenance works, and very probably, the telescope was not performing optimally 
that night. The rest of the extreme values (around day 3770) are associated with dates close to periods 
of very bad weather. The highest values of $\beta_g$ are a consequence of atmospheric-instrumental 
perturbations during the hard winter in January-February 2006. From the best solutions of $\gamma_{g, 
nm}$, the pipeline also produces the 2D inhomogeneity pattern, i.e., $\sum_{0 < n+m \leq 2} \gamma_{g, 
nm} x^n y^m$. This is depicted in Fig.~\ref{inhomomap}. In the transformation procedure, we set the 
origin of coordinates at the centre of the 1024$\times$1024 CCD, so it has an inhomogeneity correction 
equal to zero. In Fig.~\ref{inhomomap}, we see an inhomogeneity amplitude of $\sim$ 80 mmag, which is 
consistent with results from other optical telescopes \citep[e.g.,][]{Man01,Mag04}. It is evident that 
the inhomogeneity pattern in Fig.~\ref{inhomomap} plays a role in achieving 1$-$2\% photometric 
accuracy.

\begin{figure}
\centering
\includegraphics[angle=0,width=7cm]{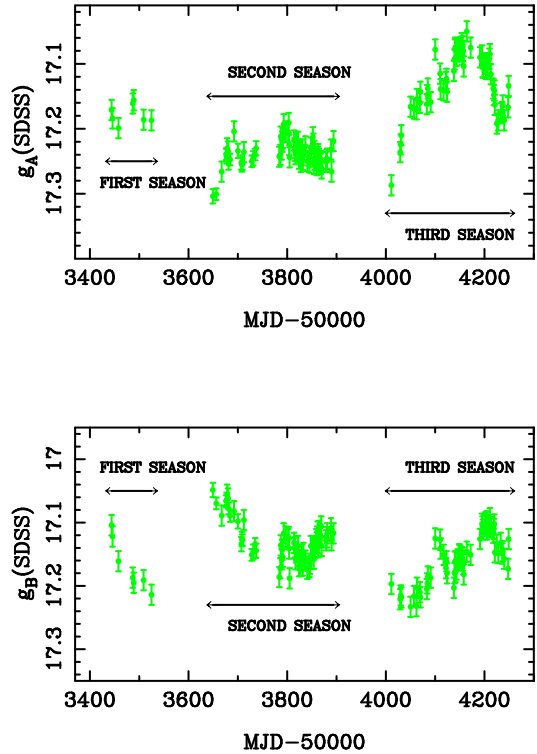}
\caption{Final magnitudes of Q0957+561A (top panel) and Q0957+561B (bottom panel) in the $g$ band of 
the SDSS photometric system. These $g$-SDSS light curves include noticeable fluctuations covering a 
2.5-year monitoring period from January 2005 to June 2007.}
\label{gmag}
\end{figure}
\begin{figure}
\centering
\includegraphics[angle=0,width=7cm]{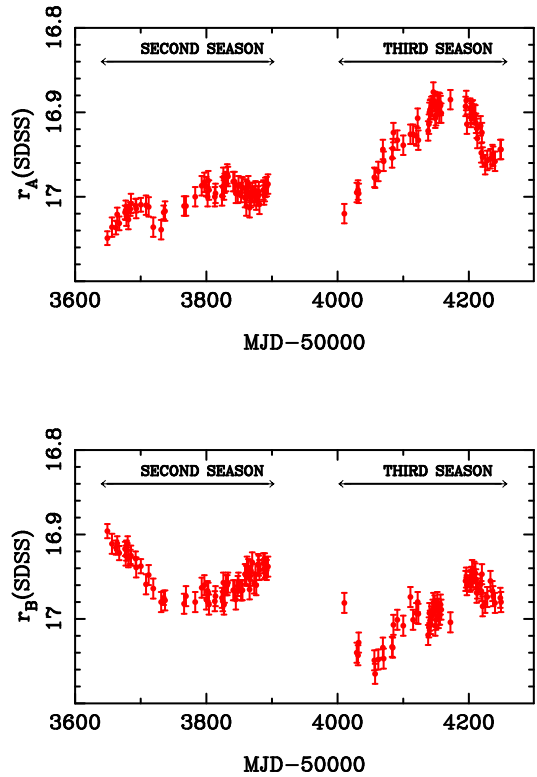}
\caption{Final magnitudes of Q0957+561A (top panel) and Q0957+561B (bottom panel) in the $r$ band of 
the SDSS photometric system. The $r$-SDSS records from October 2005 to June 2007 (two whole seasons) 
incorporate different prominent features that are also seen in the $g$-SDSS curves (see 
Fig.~\ref{gmag}), with the $g$-SDSS features having a larger amplitude.}
\label{rmag}
\end{figure}

After the $g$-band fit, the pipeline computes the calibrated and corrected records of the reference 
stars and both quasar images from Eq. (A.13). The 14$-$15th magnitude stars have a typical scatter of 
5 mmag, the 15$-$16th magnitude stars are characterized by a slightly larger scatter of about 7 mmag, 
and the faintest $\sim$ 18th magnitude star (R star) has a scatter of about 17 mmag. Although we have 
several nights with two or three exposures at different times, the standard intranight deviations of 
the stellar curves do not trace their scatters (see however Paper I). This is not surprising because 
the intranight variations exclusively correspond to several nights in the second season, which covers 
a small fraction of the total monitoring period. Thus, after some preliminary test using the stellar 
records, we find a non-biased estimator of uncertainties (typical errors): stellar scatters are well 
traced by the standard deviations between adjacent magnitudes that are separated by $\leq$ 3 d. As 
this error estimator gives reasonable results, we apply it to the $g$-SDSS magnitudes of 
\object{Q0957+561}. 

A final refinement (selection) is taken into account. Our last selection criterion is colour based: 
frames with extreme colour coefficients (see the triangles and open circles in Fig.~\ref{colour}) are 
also removed from the data set. This leads to 163 surviving frames. We obtain uncertainties (see above) 
of about 16 mmag in both $\sim$ 17th magnitude quasar components, i.e., photometry to the 1$-$2\% is 
achieved for the lensed quasar. These typical errors are in complete agreement with the stellar scatters, 
since they are clearly larger than 5$-$7 mmag (results for the brightest reference stars) and similar 
to the result for the faintest reference star R. For each component, we also group pairs or trios of 
magnitudes measured on the same night. The final light curves of \object{Q0957+561A} (top panel) and 
\object{Q0957+561B} (bottom panel) are shown in Fig.~\ref{gmag}. These $g$-SDSS light curves include 
important gradients and prominent events, which resemble those reported by \citet{Kun97} using APO 
observations in the $g$ band. The whole monitoring period consists of three observational seasons:  
January-June 2005 (first season), October 2005-June 2006 (second season), and October 2006-June 2007 
(third season). Besides the three observational seasons, there are two important gaps in the LT 
monitoring as a consequence of the annual occultation of the lens system. 

The whole procedure in the $g$ band is repeated in the $r$ band. All frames with extreme colour 
coefficients are not considered in building the final light curves, so we use a data set incorporating 142 
frames. With respect to the quasar brightness records, we achieve $\sim$ 1\% photometry (errors of about 
12 mmag). The final (grouped) magnitudes are presented in Fig.~\ref{rmag}, where the top and bottom 
panels display the records of \object{Q0957+561A} and \object{Q0957+561B}, respectively. These $r$-SDSS 
light curves\footnote{The $gr$ records are available at http://grupos.unican.es/glendama/.} trace 
prominent fluctuations that are weaker than the corresponding fluctuations in $g$-SDSS (see 
Fig.~\ref{gmag} and subsection 3.2). A similar result was claimed by the APO team \citep{Kun95,Kun97}, 
and some evidence for chromatic variability was also suggested by \citet{Ull03} \citep[see also the 
$BVRI$ variations in][]{Ser99}. The LT records in the red region of the optical spectrum are less noisy 
than previous curves at red wavelengths \citep[e.g.,][]{Kun97,Ser99}, which is due to a combination of 
an absence of relatively short variations and strict selection procedures. 


\section{Time delays of Q0957+561}

\subsection{Delay between quasar components}

The $g$-band light curve of A in the second season (October 2005-June 2006) shows significant 
fluctuations that are repeated in the $g$-band light curve of B during the third season (October 
2006-June 2007). Taking into account the expected delay range of 415$-$435 d 
\citep{Kun97,Osc01,Goi02,Ova03a}, this result is fully consistent with the presence of intrinsic 
fluctuations in those records. We use the $g$-SDSS magnitudes of A and B in the second and third 
seasons, respectively, to accurately measure the time delay(s) between both components of 
\object{Q0957+561}.

\begin{figure}
\centering
\includegraphics[angle=-90,width=10cm]{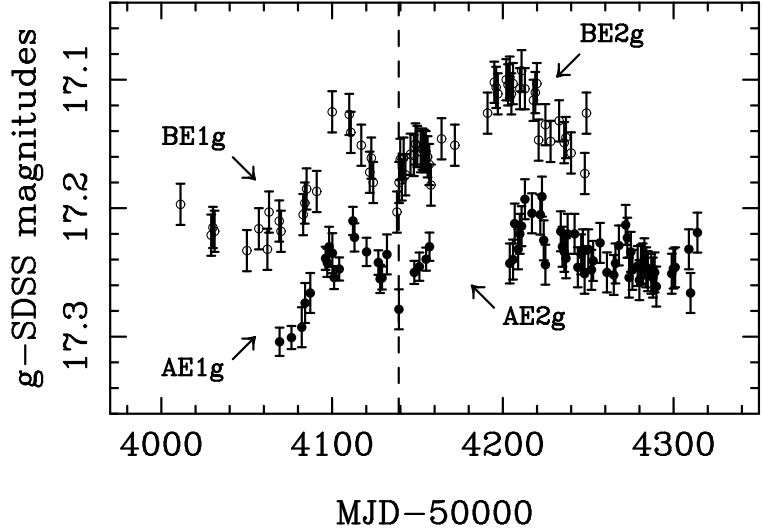}
\caption{Comparison between the $g$-band light curve of A in the second season (shifted by 420 d; see 
main text) and the $g$-band light curve of B in the third season. The A record (filled circles) shows two 
different features separated by a gap of about 50 d: while the first feature contains an event AE1$g$ and 
the beginning of another consecutive event AE2$g$, the second feature describes the (noisy) decline in 
flux of AE2$g$. A vertical line is drawn to distinguish between the two events AE1$g$ and AE2$g$. Replica 
events BE1$g$ and BE2$g$ are clearly seen in the B record (open circles).}
\label{gevents}
\end{figure}

About one half of the frames with 80 $< SNR <$ 100 correspond to the second season, and thus, they 
could help to trace the variability of A and to minimize uncertainties in time delay estimates. Their 
photometric outputs (magnitudes of A) are consistent with results from $SNR \geq$ 100 frames at 
adjacent epochs, so we recover them and expand the $g$-band record of A in the second season. 
In Fig.~\ref{gevents}, the A light curve, shifted by 420 d (filled circles), and the unchanged B light 
curve (open circles) are plotted. A reference value of 420 d is used to shift in time one component 
and to compare it with the other (see above and Introduction). The A record shows two different 
features separated by a gap of about 50 d (caused by atmospheric-instrumental problems in 
January-February 2006; see subsection 2.3). The first feature in the A curve consists of an event AE1$g$ 
and the beginning of another consecutive event AE2$g$, whereas the second feature is a noisy trend 
associated with the decline in flux of AE2$g$. These two consecutive events have an amplitude of about 
100 mmag and a duration of 50$-$150 d, and similar fluctuations BE1$g$ and BE2$g$ are evident in the B 
record. 

\begin{table}
\begin{minipage}[t]{\columnwidth}
\caption{Magnitude offset and time delay measurements in the $g$ band.}
\label{gdelay}
\centering
\renewcommand{\footnoterule}{}  
\begin{tabular}{lccc}
\hline\hline
Brightness records & Method & Offset\footnote{Magnitude offset between the A and B components, 
where the sign "$-$" means that A is fainter (see Fig.~\ref{gevents}). From $\delta^2$ we do not measure 
the shift in magnitude, since $\delta^2$ is a technique based on autocorrelation and cross-correlation 
functions. All measurements are 1$\sigma$ intervals.} (mag) & Delay\footnote{Delay of the replica 
variation in B with respect to the variation in A (the A component is leading). All measurements are 
1$\sigma$ intervals.} (d)\\                  
\hline
A(season 2)-B(season 3) &$\chi^2$   &$-$0.090 $\pm$ 0.004 &417 $\pm$ 2\\
                        &$D^2$      &$-$0.092 $\pm$ 0.004 &416 $\pm$ 5\\
AE1-BE1                 &$\chi^2$   &$-$0.083 $\pm$ 0.006 &417 $\pm$ 2\\
                        &$\delta^2$ &-                    &417 $\pm$ 2\\
\hline
\end{tabular}
\end{minipage}
\end{table}

\begin{figure}
\centering
\includegraphics[angle=0,width=10cm]{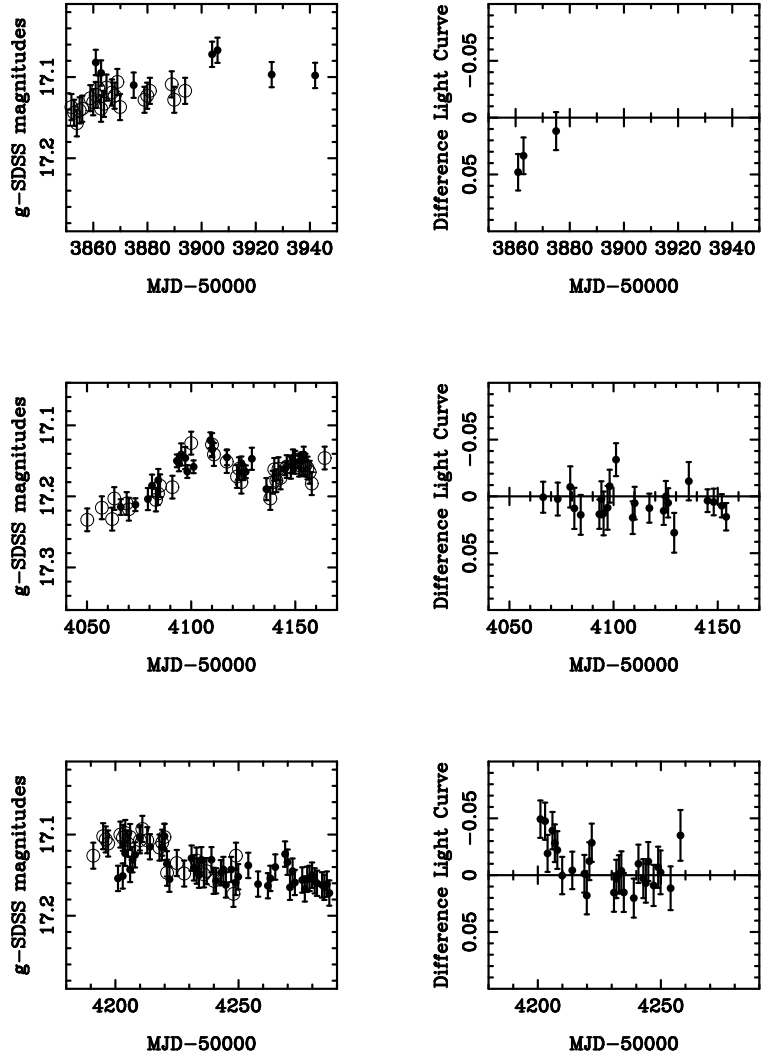}
\caption{Overlapping periods and difference light curves in the $g$ band. We show the overlap between 
the A (filled circles) and B (open circles) whole records, when the A magnitudes are shifted by the best 
solutions of the time delay and the magnitude offset (left panels). We also draw the difference light 
curve (right panels). The three overlap periods cover $\sim$ 20 d (top panels), $\sim$ 90 d (middle 
panels), and $\sim$ 60 d (bottom panels).}
\label{gdlc}
\end{figure}

Firstly, we analyse the twin events AE1$g$-BE1$g$ and AE2$g$-BE2$g$. The $S/N$ values for them (the
ratios between their semi-amplitudes and their mean photometric errors) are $(S/N)_{AE1g} \sim$ 4 and 
$(S/N)_{BE1g} \sim (S/N)_{BE2g} \sim$ 3.4. In spite of the fact that AE2$g$ is a prominent event, it is 
poorly traced as a consequence of the 50-day gap and the noisy right wing. Thus, we are not able to 
determine a reliable value of $(S/N)_{AE2g}$, and the effective signal-to-noise ratio for AE2$g$ could 
be significantly less than 3$-$4. The difficulties in inferring a time delay from the AE2$g$-BE2$g$ 
twin events confirm our suspicions. Unfortunately, it is not possible to measure two independent delays, 
one from AE1$g$-BE1$g$ and the other from AE2$g$-BE2$g$. The only options are the estimation of a 
delay related to the two flares in the source of intrinsic variability, i.e., using all events in 
Fig.~\ref{gevents}, or a delay corresponding to the first flare, i.e., from AE1$g$-BE1$g$. 

Secondly, to calculate the two-flare time delay and magnitude offset (i.e., a constant magnitude 
shift between the light curves of the two quasar components), we use two techniques: $\chi^2$ 
minimization \citep[e.g.,][]{Kun97,Ull06} and the minimum dispersion ($D^2$) method \citep{Pel94,Pel96}, 
characterized by a bin semisize ($\alpha$) and a decorrelation length ($\delta$). The 
choice of $\alpha = \delta$ = 9 d is a good compromise between the A-B connection and time resolution. 
Through the $\chi^2$ minimization ($\alpha$ = 9 d), we obtain the best solutions of the delay and magnitude
offset: $\Delta t_{BA}$ = 417 d and $\Delta m_{BA}$ = $-$0.090 mag ($\chi^2 \sim$ 1.2).The sign "$-$" 
in the $\Delta m_{BA}$ value means that the A component is fainter. The $D^2$ minimization ($\delta$ = 9 
d) gives the best solutions of $\Delta t_{BA}$ = 416 d and $\Delta m_{BA}$ = $-$0.092 mag. The uncertainties 
in the magnitude offset and time delay are inferred from 1000 repetitions of the experiment (synthetic 
light curves based on the observed records). The 1$\sigma$ intervals appear in Table~\ref{gdelay}. 
Table~\ref{gdelay} indicates that the error in time delay from the $\chi^2$ minimization is 
substantially less than the error from the minimum dispersion method. Both measurements of the 
two-flare time delay are consistent with the APO main delay in the $g$ band (see Introduction). 

Thirdly, we exclusively use the AE1$g$-BE1$g$ twin events. The idea is to accurately measure the 
gravitational lens delay associated with only one flare produced in the source of variability. This 
time we focus on the $\delta^2$ method (see, e.g., Paper I) and the $\chi^2$ minimization, which 
produces a delay error smaller than the delay uncertainty from the minimum dispersion technique (see 
Table~\ref{gdelay}). The $\delta^2$ technique obtains the optimal match between the time-shifted 
discrete autocorrelation function ($DAF$) and the discrete cross-correlation function 
\citep[$DCF$;][]{Ede88}. From the $\chi^2$ minimization ($\alpha$ = 9 d), the best solutions of the time 
delay and magnitude offset are 417 d and $-$0.083 mag, respectively ($\chi^2 \sim$ 0.9). From the 
$\delta^2$ method and 1000 synthetic light curves, the delay measurement (1$\sigma$ interval) is 
identical to that derived from the $\chi^2$ technique (see Table~\ref{gdelay}). Therefore, the LT first 
twin events are useful to determine a robust time delay $\Delta t_{BA}$ = 417 $\pm$ 2 d (1$\sigma$). 
This is fully consistent with the APO main delay \citep{Kun97,Goi02}. The $\delta^2$ analysis 
also indicates that $\Delta t_{BA} \leq$ 424 d (99\% confidence interval), so the AE1$g$-BE1$g$ delay is 
inconsistent (at about the 3$\sigma$ level) with the APO secondary delay \citep{Goi02}. The 
$r$-band curves of AE1-BE1 are not used to determine a time delay because $S/N <$ 3 for these twin 
events \citep[e.g.,][]{Pij97}. 

We now check for the possible existence of extrinsic variability in our records. We compute the 
difference light curve between the A and B components, since no extrinsic variability should result in 
a flat difference light curve. To obtain the difference light curve, the magnitude- and time-shifted 
light curve of image A is subtracted from the light curve of image B \citep[e.g.,][]{Sch98,Gil01}. In 
Fig.~\ref{gdlc} (left panels), we show the overlap between the A (filled circles) and B (open circles) 
whole records, when the A magnitudes are shifted by the best solutions of the time delay and the 
magnitude offset. The difference light curve is also plotted in the right panels of Fig.~\ref{gdlc}. 
The overlap between A-first season and B-second season covers a very short period of about 20 d (see 
the top panels of Fig.~\ref{gdlc}). For this overlap period, the difference curve contains two 
consecutive deviations from the zero level, which are not significative \citep[e.g.,][]{Gil01}. The 
overlap between A-second season and B-third season is much more important than the first overlap 
(in the top panels). In the middle panels of Fig.~\ref{gdlc}, we display the situation before the 
50-day gap (see above and Fig.~\ref{gevents}), where the difference curve has a noisy trend that is 
consistent with zero. In the bottom panels, the behaviour after the 50-day gap is shown. In this 
last period, the difference curve is also mainly noise. However, a clear event appears at the 
beginning of the overlapping period, i.e., six consecutive points are placed above the zero level. 
Although this naively could be interpreted as the wing of a microlensing event (i.e., extrinsic 
variability), the A data were obtained at the end of a hard winter in which the colour coefficient 
strongly deviated (40$-$70\%) from its average value. While some frames with extreme colour 
coefficients are not considered in the analysis (see the triangles and the open circle around day 3770 
in Fig.~\ref{colour}), additional adjacent frames are also unsuitable for fine variability studies. 
Therefore, bad weather and anomalous behaviour of the LT devices are the most reasonable explanations 
for the anomalous variation in A that is simultaneously observed in both components. In summary, we do 
not find evidence of extrinsic variability in the light curves of \object{Q0957+561}.

\subsection{Delay between optical bands}

\begin{figure}
\centering
\includegraphics[angle=0,width=10cm]{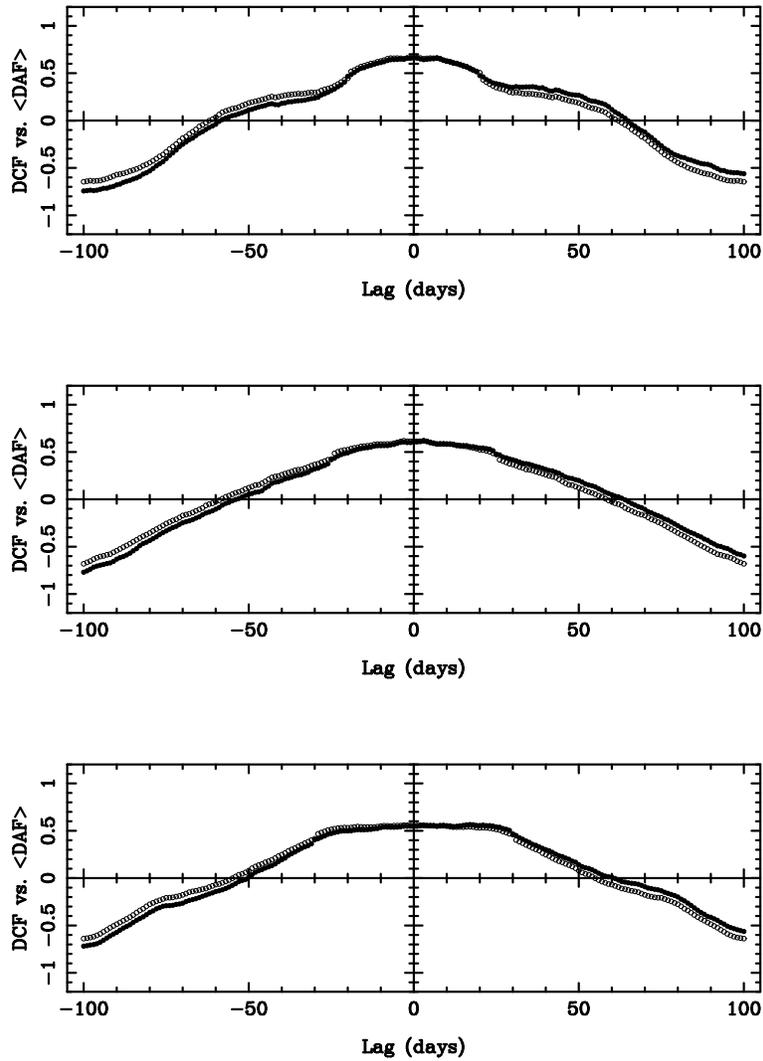}
\caption{Comparison between the $DCF$ (filled circles) and the $\langle DAF \rangle$ (open circles). While
the $DCF$ is the $gr$ cross-correlation function, the $\langle DAF \rangle$ is the average of the $gg$ and $rr$ 
autocorrelation functions. We use the AE3$g$-AE3$r$ events (see main text) and three bin semisizes: $\alpha$ =
20 (top panel), 25 (middle panel), and 30 (bottom panel) d.}
\label{grcorr}
\end{figure}
\begin{figure}
\centering
\includegraphics[angle=-90,width=10cm]{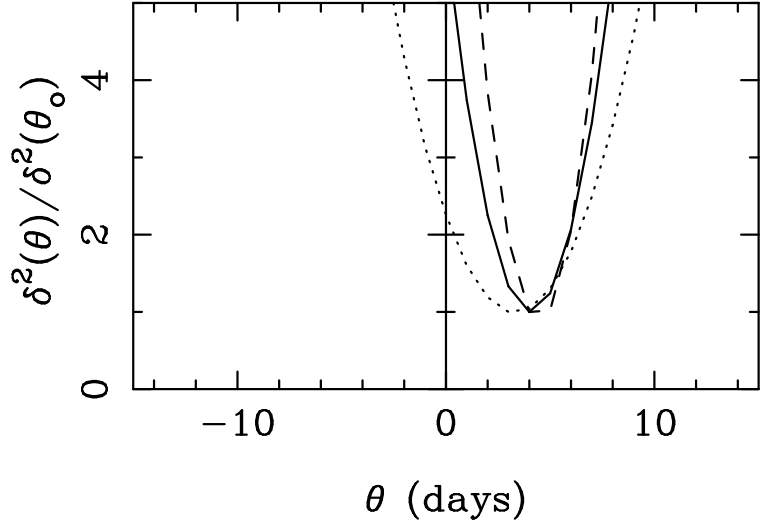}
\caption{Normalised $\delta^2$ function from the AE3$g$-AE3$r$ events. We use bin semisizes $\alpha$ = 20 
(dotted line), 25 (dashed line), and 30 (solid line) d. In Fig.~\ref{grcorr}, we can observe the presence of 
time shifts between the $DCF$ and $\langle DAF \rangle$, which translate into interband delay peaks centered 
on 3$-$4.5 d (AE3$g$ leading AE3$r$).}
\label{grdelta}
\end{figure}

The time delay between optical-UV continuum flux variations at two different wavelengths can be used to 
test the variability secenario \citep[e.g.,][]{Col99}. It might be produced by reprocessing of high 
energy radiation in an accretion disc around a supermassive black hole. The reprocessing hypothesis 
assumes that the optical-UV variations are the response of the gas in the disc to higher-energy 
fluctuations in the vecinity of the disc axis. Moreover, the existence of a radiative coupling between 
the variations is also assumed, i.e., the time delay represents a light-travel time between two disc 
annuli \citep[see details in][]{Col99}. \citet{Col99} measured two time lags between fluctuations at 
two optical wavelengths and the corresponding UV fluctuations (UV variability leading optical variations) 
in the records of \object{NGC 7469} at $z$ = 0.016, and they found a good agreement between their delay 
estimates ($\sim$ 1$-$2 d) and reverberation within an accretion disc. \citet{Ser05} and 
\citet{Cac07} also explored the thermal reprocessing hypothesis in local active galactic nuclei. For 
\object{Q0957+561}, \citet{Col01} reported a delay of about 3.4 d between the $r$-band and $g$-band APO 
main events ($g$-band events leading those in the $r$ band), which translates into a rest-frame lag of 
about 1.4 d, in excellent agreement with predictions of the disc reprocessing scenario. This first delay 
between optical bands for a GLQ requires an independent confirmation as well as new efforts with other 
GLQs \citep[e.g.,][]{Kop06}, and here we try to reach the first goal.  

For such a task, we focus on the LT events with highest $S/N$. The AE1-BE1 twin events are ruled out 
because $(S/N) <$ 3 in the $r$ band. However, there are two very prominent variations around day 4150 in 
the top panels of Figs.~\ref{gmag}$-$\ref{rmag} (A component). These AE3$g$ and AE3$r$ variations last 
$\sim$ 250 d (the whole light curves of A in the third season are considered as large events) and have 
signal-to-noise ratios above 6. We use fluxes in arbitrary units $f = 10^7 \times 10^{-0.4m}$ to compare 
AE3$g$ and AE3$r$. The use of fluxes (instead of magnitudes $m$) permits a fair cross-correlation between 
two records that, apart from a possible delay, differ in a multiplicative constant and an additive 
constant. On average, the light curves were sampled two times per week. However, there are 20-day 
gaps around day 4180. Unfortunately, due to a combination of the kind of variability (time asymmetric 
events consisting of slow rises and rapid declines) and these short gaps close to the maxima, it is not 
possible to infer a reliable $DCF$ with good time resolution, i.e., $\alpha \leq$ 10 d (see above). For 
$\alpha <$ 20 d, 20-day artifacts at lags of $\pm$ 50 d appear in the $DCF$. This unphysical signal at 
$\pm$ 50 d is only avoided using longer bins with $\alpha \geq$ 20 d, so we are forced to take relatively 
long bins. This is not a problem at all, but the measurement would be more accurate (but not more reliable) 
with better time resolution. Some $DCF$ (filled circles) and $\langle DAF \rangle$ (open circles) trends 
are shown in Fig.~\ref{grcorr}. The top, middle, and bottom panels of Fig.~\ref{grcorr} contain the 
results for $\alpha$ = 20, 25, and 30 d, respectively. Here, $\langle DAF \rangle$ is the average of the 
$gg$ and $rr$ autocorrelation functions, whereas $DCF$ represents the $gr$ cross-correlation function. 


\begin{table}
\begin{minipage}[t]{\columnwidth}
\caption{Time lag measurements from the AE3$g$-AE3$r$ events.}
\label{grdelay}
\centering
\renewcommand{\footnoterule}{}  
\begin{tabular}{ccc}
\hline\hline
$\alpha$\footnote{We use the $\delta^2$ technique (see main text) and five values of the bin semisize 
$\alpha$.} (d) & Time lag\footnote{All measurements are 1$\sigma$ intervals, and positive lags mean that 
the $r$-band event is delayed in relation to the arrival of the associated $g$-band event.} (d) & 
Probability of lags $\leq$ 0 (\%)\\                  
\hline
20 &3.0 $\pm$ 2.0 &11.6\\
25 &4.5 $\pm$ 2.5 &6.9\\
30 &4.0 $\pm$ 2.0 &7.5\\
35 &3.5 $\pm$ 2.0 &8.5\\
40 &3.5 $\pm$ 2.0 &9.3\\
\hline
\end{tabular}
\end{minipage}
\end{table}

In Fig.~\ref{grcorr}, there are no important distortions in the features of the $DCF$ compared to 
the features in the $\langle DAF \rangle$, but the existence of a delay of several days is evident. In 
other words, to get an optimal match, the $\langle DAF \rangle$ should be shifted by several days. Possible 
values of this time shift ($\theta$) versus the associated $\delta^2(\theta)$ values normalised by its 
minimum value $\delta^2(\theta_0)$ are plotted in Fig.~\ref{grdelta}. The $\delta^2(\theta)$ function was 
defined in Eq. (7) of \citet{Ser99} (see also above), and we use $\alpha$ = 20 (dotted line), 25 (dashed 
line), and 30 (solid line) d in Fig.~\ref{grdelta}. This figure displays relatively narrow peaks centered 
on 3$-$4.5 d (best values of the interband delay; AE3$g$ leading AE3$r$). Uncertainties are again computed 
by applying the $\delta^2$ minimization to 1000 synthetic data sets. Through the distributions of delays 
($\alpha$ = 20$-$40 d), five 1$\sigma$ measurements are presented in Table~\ref{grdelay}. The $\delta^2$ 
results in Table~\ref{grdelay} agree with the previous time lag determination from APO light curves, and 
we adopt $\Delta t_{rg}$ = 4.0 $\pm$ 2.0 d (using an intermediate bin semisize $\alpha$ = 30 d; the 
probability of $\Delta t_{rg} \leq$ 0 is only 7.5\%) as our final 1$\sigma$ measurement. 


\section{Summary and conclusions}  

Liverpool Quasar Lens Monitoring is a long-term project to follow the optical ($gri$ bands) variability 
of 10$-$20 GLQs with the Liverpool robotic telescope \citep{Ste04}. The first phase of this project (LQLM 
I) was conducted between January 2005 and July 2007. While in Paper I we mainly studied the intrinsic 
variability of \object{Q0909+532} in the $r$ band, in this paper we present the monitoring programme of 
\object{Q0957+561} in the $gr$ bands. A main goal of our project (LQLM) is to considerably increase the 
public database of GLQs. Thus, all LQLM I pre-processed frames of \object{Q0909+532} and 
\object{Q0957+561} are publicly available on the Lens Image Archive of the German Astrophysical Virtual 
Observatory\footnote{See the Web site http://vo.uni-hd.de/lensdemo/view/q/form.}. 

We have fully developed two photometric pipelines through the 3 years of observations and analyses. The 
transformation pipeline incorporates zero-point, colour, and inhomogeneity corrections in the 
instrumental fluxes, so photometry to the 1$-$2\% is achieved for \object{Q0957+561A} and 
\object{Q0957+561B}. We detect an inhomogeneous response over the CCD area, which has an amplitude of 
$\sim$ 80 mmag (from maximum to minimum) and is consistent with studies in other optical telescopes 
\citep[e.g.,][]{Man01,Mag04}. Moreover, the colour coefficient is allowed to vary through time, because 
the atmospheric-instrumental conditions signicantly evolve through 2.5 years of monitoring. Due to 
atmospheric-instrumental problems at some epochs, the colour coefficient reaches anomalous values, i.e., 
we obtain dramatic deviations with respect to the average coefficient. Thus, the frames corresponding to 
an anomalous coefficient are removed or not considered. 

The LT $gr$ light curves of \object{Q0957+561} show several prominent events and gradients, and some 
of them (in the $g$ band) are used to infer a time delay between components $\Delta t_{BA}$ = 417 $\pm$ 2 
d (1$\sigma$). This gravitational lens delay from new $g$-band events is in agreement with the delay 
from the previous APO $g$-band main events \citep{Kun97}, so the associated UV flares in the variable 
source (APO and LT events) probably originate in the same emission region \citep{Yon99}. Taking into 
account that the previous APO $gr$-band main events are plausibly due to reverberation within an 
irradiated accretion disc \citep{Col01}, the new $gr$-band events are likely related to flares in the 
central accretion disc. In addition, the delay between the two new LT large events in the $g$ and $r$ bands: 
$\Delta t_{rg}$ = 4.0 $\pm$ 2.0 d (1$\sigma$; the $g$-band event is leading), coincides with the 
estimation by \citet{Col01} and agrees with flares generated during reprocessing in the accretion disc. 
Therefore, most APO-LT variations in the $g$ and $r$ bands are very probably associated with the gas disc 
around the supermassive black hole (only the APO secondary events have been associated with flares that 
were produced far away from the accretion disc; see Introduction). The detection of the same interband 
delay (between the $g$ and $r$ band) in the two monitoring campaigns (APO and LT) also suggests that the 
accretion disc reprocessing in \object{Q0957+561} is a usual occurance at different times for 
different prominent flares. Hence, very likely, most observed variations in the $g$ and $r$ bands (APO and 
LT fluctuations with an amplitude of $\sim$ 100 mmag and lasting $\sim$ 100 d) are associated with 
reverberation within the gas disc around the supermassive black hole. 

We add 2.5 years of time coverage to the previous 1.5-year $gr$-band records of \object{Q0957+561}, 
and remark that our difference light curves are consistent with zero. Thus, there is no evidence of 
extrinsic variations in both APO and LT independent experiments separated by $\sim$ 10 years. These 
results disagree with the claim by \citet{Sch05} that microlenses in the lensing galaxy affect the
observed variability. Therefore, the complex quasar structure suggested by this author is not supported
by the $gr$-band light curves of \object{Q0957+561}. We also remark that double replicas in the records of
the B component are not detected in the APO and LT experiments. This clearly contradicts previous 
conclusions by \citet{Hir07}, which indicated that the full B light curve of the lensed quasar can be
decomposed into a sum of two similar and time shifted curves. Finally, the APO-LT combined database of 
\object{Q0957+561} (together with other monitorings done between both experiments) is a promising tool for 
studying the quasar structure and the composition of the lensing halo 
\citep[e.g.,][]{Sch98,Koc04b}.

\begin{acknowledgements}
We thank an anonymous referee for several comments that improved the presentation of our results. 
We also thank C. Moss for guidance in the preparation of the robotic monitoring project with the 
Liverpool telescope. The Liverpool Telescope is operated on the island of La Palma by Liverpool 
John Moores University in the Spanish Observatorio del Roque de los Muchachos of the Instituto 
de Astrofisica de Canarias with financial support from the UK Science and Technology Facilities 
Council. We thank B. McLeod for providing the IMFITFITS software to us. We use information taken 
from the Sloan Digital Sky Survey (SDSS) Web site, and we are grateful to the SDSS team for doing 
that public database. This research has been supported by the Spanish Department of Education and 
Science grants AYA2004-08243-C03-02 and AYA2007-67342-C03-02, University of Cantabria funds, grant 
for young scientists of the President of the Russian Federation (number MK-2637.2006.2), Deutscher 
Akademischer Austausch Dienst (DAAD) grant number A/05/56557 and grant of Russian Foundation for 
Basic Research (RFBR) 06-02-16857. EK holds the Taiwan National Research Councils grant No. 
96-2811-M-008-058. RGM holds a grant of the ESP2006-13608-C02-01 project financed by the Spanish
Department of Science and Innovation.
\end{acknowledgements}

\appendix

\section{Transformation Equations}

The initial transformation equations for a given reference star are
\begin{eqnarray}
g^*(t_j) & = & g + A_g(t_j) + C_g(t_j)(g - r)   , \\
r^*(t_k) & = & r + A_r(t_k) + C_r(t_k)(r - i)   , 
\end{eqnarray}
where $g^*$ and $r^*$ are the instrumental magnitudes of the star, $g$, $r$, and $i$ are its 
standard magnitudes, $A_g$ and $A_r$ are the zero-point terms (including instrumental 
and atmospheric corrections), and $C_g$ and $C_r$ are the colour coefficients. The zero-point 
terms and the colour coefficients are expected to significantly change during the 2.5-year 
monitoring period, so we explicitly consider their time evolution. Here, $t_j$ and $t_k$ denote 
observation times in the $g$ and $r$ bands, respectively. As we are initially interested in 
the usual systematic corrections, Eqs. (A.1$-$2) do not include other possible terms (see here below). 
Instead of the LT photometric system ($ugriz \equiv u'g'r'i'z'$), we want to use the SDSS 
"natural" system, since accurate standard magnitudes are available in this SDSS 2.5m system 
\citep[e.g.,][]{Smi02,Sto02}. SDSS magnitudes are also suitable for comparing our results with 
future data of \object{Q0957+561} using different facilities and/or SDSS quasar studies/databases 
\citep[e.g.,][]{Van04,Sch07}. From equations for transforming LT magnitudes to magnitudes in the 
SDSS system\footnote{See the Web site http://www.sdss.org/dr6/algorithms/.}:
\begin{eqnarray}
g & = & g_{SDSS} + B_g(g - r) + K_g    , \\
r & = & r_{SDSS} + B_r(r - i) + K_r    ,      
\end{eqnarray}
and equations that relate LT and SDSS colours:
\begin{eqnarray}
g - r & = & a_{gr}(g - r)_{SDSS} + b_{gr}    , \\
r - i & = & a_{ri}(r - i)_{SDSS} + b_{ri}    ,           
\end{eqnarray}
it is possible to rewrite Eqs. (A.1$-$2) as
\begin{eqnarray}
g^*(t_j) & = & g_{SDSS} + \alpha_g(t_j) + \beta_g(t_j)(g - r)_{SDSS}  , \\
r^*(t_k) & = & r_{SDSS} + \alpha_r(t_k) + \beta_r(t_k)(r - i)_{SDSS}  . 
\end{eqnarray}
The $\alpha_g$ term and the $\beta_g$ coefficient are given by (it is trivial to write  
expressions for $\alpha_r$ and $\beta_r$) 
\begin{eqnarray}
\alpha_g(t_j) & = & A_g(t_j)  + K_g + b_{gr}[B_g + C_g(t_j)]    , \\
\beta_g(t_j) & = & a_{gr}[B_g + C_g(t_j)]    . 
\end{eqnarray}
Taking into account typical values of $C_g$ ($\sim -0.029$) and $C_r$ ($\sim 0.034$) reported by
the LT team (on the LT Web site; see main text), the SDSS estimates of $B_g$ ($\sim -0.060$) and 
$B_r$ ($\sim -0.035$) on the SDSS Web site (see here above), and $a_{gr} \sim a_{ri} \sim 1$, we
expect typical colour coefficients $\beta_g \sim -0.089$ and $\beta_r \sim -0.001$. On the other
hand, the adopted standard magnitudes of the reference stars (XGFHDER field stars; see main text) 
appear in Table~\ref{SDSSstars}. These are PSF magnitudes in the SDSS catalogue$^2$.

In order to achieve 1$-$2\% photometric accuracy with the RATCam camera (on the LT), one additional
detail must be taken into account in the transformation equations (A.7$-$8). We introduce an 
inhomogeneity term that corrects the flat-field systematic error over the camera area, which might 
have a total amplitude of $\sim$ 50 mmag \citep[e.g.,][]{Man01,Mag04}. For example, this kind of 
error could be related to twilight flats. During twilight exposures, some scattered light (within 
the dome) would reach the camera, and thus, the illumination would not be homogeneous. This effect
invalidates the basic hypothesis of homogeneous illumination. Here, we assume a second order 2D 
polynomial to account for the inhomogeneity term, so the final transformation equations are 
\begin{eqnarray}
g^*(t_j) & = & g_{SDSS} + \alpha_g(t_j) + \beta_g(t_j)(g - r)_{SDSS} + 
\sum_{0 < n+m \leq 2} \gamma_{g, nm} x^n(t_j) y^m(t_j)    , \\
r^*(t_k) & = & r_{SDSS} + \alpha_r(t_k) + \beta_r(t_k)(r - i)_{SDSS} + 
\sum_{0 < n+m \leq 2} \gamma_{r, nm} x^n(t_k) y^m(t_k)    , 
\end{eqnarray}
where ($x$,$y$) is the 2D position of the star on the CCD. To find the relevant parameters in the 
$g$ band, i.e., $\alpha_g(t_j)$, $\beta_g(t_j)$, and $\gamma_{g, nm}$, we may fit the observed 
magnitude deviations (instrumental $-$ standard) of the seven reference stars to the model 
incorporating the three systematic terms: zero-point, colour, and inhomogeneity. Once the fit has 
been made, the $g$-SDSS magnitude of any point-like source (star or quasar) is derived in a 
straightforward way:
\begin{equation}
g(SDSS) = g_{SDSS} + \delta = g^*(t_j) - \alpha_g(t_j) - \beta_g(t_j)(g - r)_{SDSS} - 
\sum_{0 < n+m \leq 2} \gamma_{g, nm} x^n(t_j) y^m(t_j)    .
\end{equation}
In Eq. (A.13), $\delta$ represents the deviation caused by random noise (e.g., photon noise) and 
unkown (but presumibly small) systematic corrections. For a non-variable star (e.g., a reference 
star), variations in $g(SDSS)$ are generated by noise ($\delta$). However, for variable stars or 
quasars, there are two kinds of variability. While true variability is due to time evolution of 
$g_{SDSS}$, noise is an additional source of fluctuations. The $(g - r)_{SDSS}$ colour of 
\object{Q0957+561} might also evolve over time. Thus, the use of an average colour $\langle (g - 
r)_{SDSS} \rangle$ in the colour correction introduces a systematic noise $\delta_{col} = 
\beta_g(t_j)\delta(g - r)_{SDSS}$ associated with the colour variation. Fortunately, for moderate 
fluctuations with amplitude of $\sim$ 25 mmag \citep[e.g.,][]{Kun95}, the amplitude of the colour 
noise is only $\sim$ 0.2\%, well below our accuracy goal. The $r$-SDSS magnitude of a source is 
given by an expression similar to Eq. (A.13).

\begin{table}
\begin{minipage}[t]{\columnwidth}
\caption{Adopted standard magnitudes of the reference stars.}
\label{SDSSstars}
\centering
\renewcommand{\footnoterule}{}  
\begin{tabular}{lccc}
\hline\hline
Star & $g_{SDSS}$ & $r_{SDSS}$ & $i_{SDSS}$\\                  
\hline
X &14.213 &13.849 &13.750\\
G &14.461 &14.157 &14.060\\
F &14.513 &14.186 &14.089\\
H &15.116 &14.422 &14.174\\
D &15.485 &14.951\footnote{The SDSS catalogue seems to contain a wrong value of the $r$-SDSS 
magnitude of the D star ($r_{SDSS}$ = 15.674), so the D star would be fainter than the E star in 
this band. This disagrees with our current LT observations and several previous observations in 
the red region of the optical spectrum. Thus, the $r$-SDSS magnitude is inferred through the 
$r_{SDSS}$ vs. $VR$ relationship: $r_{SDSS} = V - 0.89(V - R) + 0.39$. This law is based on the 
$r$-SDSS magnitudes of the rest of stars and the corresponding $VR$ magnitudes in Tables 1-2 of 
\citet{Ova03b}.} &14.770\\
E &15.816 &15.217 &15.018\\
R &17.879 &16.801 &16.419\\
\hline
\end{tabular}
\end{minipage}
\end{table}

\end{document}